# PID Parameters Optimization by Using Genetic Algorithm

Andri Mirzal, Shinichiro Yoshii, Masashi Furukawa

Graduate School of Information Science and Technology
Hokkaido University
Sapporo, Japan
Email: andri, yoshii, mack@complex.eng.hokudai.ac.jp

## ABSTRACT

Time delays are components that make time-lag in systems response. They arise in physical, chemical, biological and economic systems, as well as in the process of measurement and computation. In this work, we implement Genetic Algorithm (GA) in determining PID controller parameters to compensate the delay in First Order Lag plus Time Delay (FOLPD) and compare the results with Iterative Method and Ziegler-Nichols rule results.

**Keywords**: time delay, genetic algorithm, iterative method, ziegler-nichols rule

## 1. INTRODUCTION

In the previous work [1], authors have implemented and compared two tuning methods, *Iterative Method* and *Ziegler-Nichols rule*, to compensate the effect of delay in stability of systems and showed that Iterative Method has superior performance in analyzed cases, FOLPD (First Order Lag plus Time Delay). But there are some cases where we can't use these two tuning methods, i.e. the dynamic plants which its parameters are constantly changing. In this sort of systems, we have to do retuning in real time, which can't be applied by the tuning methods because we have to take the systems offline first in order to set its parameters.

In this work, we extend our previous work [1] by implementing *Genetic Algorithm* (GA) in determining PID Controller parameters to compensate the delay in order one (FOLPD) and compare the results with Iterative Method and Ziegler-Nichols rule results.

**Figure 1**. The PID controller general structure where plant has delay component

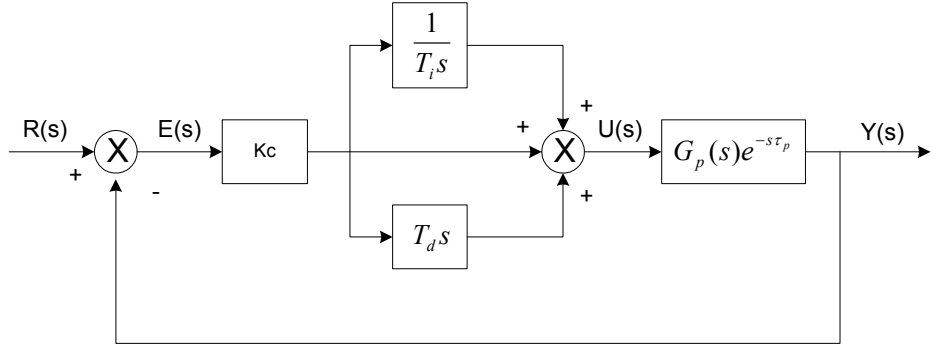

## 2. THE OBJECTIVE FUNCTIONS FITNESS VALUES

The most crucial step in applying GA is to choose the objective functions that are used to evaluate fitness of each chromosome. Some works [3] [4] use *performance indices* as the objective functions. In [3] author uses Mean of the Squared Error (MSE), Integral of Time multiplied by Absolute Error (ITAE), Integral of Absolute Magnitude of the Error (IAE), and Integral of the Squared Error (ISE), while in [4] authors use ISE, IAE, and ITAE. Here we use all four performance indices stated above and Integral of Time multiplied by the Squared Error (ITSE) to minimize the error signal $E(s)$ and compare them to find the most suitable one. The performance indices are defined as follow [2]:

$$MSE = \frac{1}{t}\int_0^\tau (e(t))^2 dt, \; ITAE = \int_0^\tau t|e(t)|dt, \; IAE = \int_0^\tau |e(t)|dt \; ISE = \int_0^\tau e(t)^2 dt, \text{ and } ITSE = \int_0^\tau te(t)^2 dt \quad (1)$$

Where $e(t)$ is the error signal in time domain.

The PID controller is used to minimize the error signals, or we can define more rigorously, in the term of error criteria: to minimize the value of performance indices mentioned above. And because the smaller the value of performance indices of the corresponding chromosomes the fitter the chromosomes will be, and vice versa, we define the fitness of the chromosomes as:

$$fitness\ value = \frac{1}{performance\ index} \quad (2)$$

## 3. DELAY COMPONENT

Delay in control systems can be defined as time-interval between an event that start in one point with its output in another point within systems [5]. Delay is also recognized as transport lag, deadtime, and time lag. Because delay always reduces stability of minimum phase systems (systems which don't have poles and zeros in the right half of s-plane), it is important to analysis stability of systems with time delay.

**Figure 2**. Delay effect on system

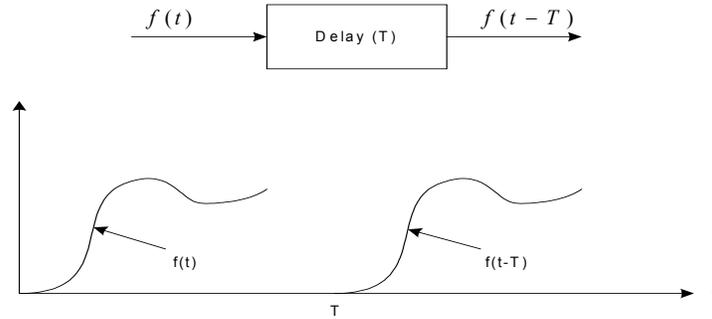

We can see delay effect in a system which causes time shift at system's output from figure above. The relationship between $f(t)$ and $f(t-T)$ can be written as:

$$\ell[f(t-T)u(t-T)] = \int_0^\infty f(t-T)u(t-T)e^{-st}dt \qquad (3)$$

where $u(t)$ is unit step. Let $\tau = t-T$,

$$\int_0^\infty f(t-T)u(t-T)e^{-st}dt = \int_{-T}^\infty f(\tau)u(\tau)e^{-s(\tau+T)}d\tau \qquad (4)$$

assume $f(t) = 0$ for $t < 0$,

$$\begin{aligned}
\int_{-T}^\infty f(\tau)u(\tau)e^{-s(\tau+T)}d\tau &= \int_0^\infty f(\tau)u(\tau)e^{-s(\tau+T)}d\tau \\
&= \int_0^\infty f(\tau)e^{-s\tau}e^{-Ts}d\tau \\
&= e^{-Ts}\int_0^\infty f(\tau)e^{-s\tau}d\tau \\
&= e^{-Ts}F(s)
\end{aligned} \qquad (5)$$

So, we get :

$$\ell[f(t-T)u(t-T)] = e^{-Ts}F(s) = e^{-Ts}\ell[f(t)u(t)] \qquad (6)$$

In order to do tuning processes using GA, we approximate the delay with Direct Frequency Response (DFR) series. Actually in Matlab, there is time delay built in function, Pade series approximation, but we choose to use DFR series because first, it has been shown in [1] that this series has the smallest average error among the others seven series and the second is while the delay block function in Control Systems Toolbox (used to simulate Iterative Method and Ziegler-Nichols tuning rules) uses Pade series, the delay component modeled by tf.m function (used to construct transfer function of a system) is still in the complex frequency domain representation (see equation 10), so that we have to translate it into polynomial series representation.

Furthermore, for simulation purpose, we use second order DFR series to circumvent unnecessary complexity, because as the order of the series are getting higher, not only the calculation becomes difficult but also it introduces new poles and zeros which make the system much more elusive.

$$\text{DFR series}: \ e^{-s\tau} \approx \frac{1 - 0.49s\tau + 0.0954s^2\tau^2}{1 + 0.49s\tau + 0.0954s^2\tau^2} \qquad (7)$$

## 4. GENETIC ALGORITHM

PID controller parameters will be optimized by applying GA. Here we use Matlab Genetic Algorithm Toolbox [6] to simulate it. The first and the most crucial step is to encoding the problem into suitable GA chromosomes and then construct the population. Some works recommend 20 to 100 chromosomes in one population. The more the chromosomes number, the better the chance to get the optimal results. However, because we have to consider the execution time, we use 80 or 100 chromosomes in each generation.

Encoding is done in real number rather than binary encoding because the latter discards the parameters value if it exceeds its precision capability. Each chromosome comprises of three parameters, $K_d$, $K_p$, $K_i$, with value bounds varied depend on the delay and objective functions used. After many experiments, we find that the value bounds should be set according to the Iterative Method and Ziegler-Nichols rule value range to ensure the convergence (there are many cases which the convergence can't be reached if we set the parameter value bounds arbitrarily, even though the optimal results included in those bounds range).

The population in each generation is represented by 80 x 4 or 100 x 4 matrix, depends on the chromosomes number in population, which each row is one chromosome that comprise $K_d$, $K_p$, $K_i$ values and the last column added to accommodate fitness values (F) of corresponding chromosomes.

$$\begin{bmatrix} K_{d1} & K_{p1} & K_{i1} & F_1 \\ K_{d2} & K_{p2} & K_{i1} & F_2 \\ . & . & . & . \\ . & . & . & . \\ K_{dn} & K_{pn} & K_{in} & F_n \end{bmatrix} \begin{matrix} \text{chromosome 1} \\ \text{chromosome 2} \\ . \\ . \\ \text{chromosome } n \end{matrix} \qquad (8)$$

We use maximum generation termination (maxGenTerm.m) to terminate the program rather than considering the best chromosome fitness values changing rate because we want to control the execution time. However, the best chromosome fitness values changing rate is also being considered by running the programs until the best fitness value stop increasing, then we set that point as the maximum generation. After several experiments it's shown that there is no visible improvement after $300^{th}$ generation, so we set 300 as the maximum generation.

Matlab GA Toolbox [6] provides three selection techniques, Tournament Selection, Roulette Wheel Selection and Normalize Geometric Selection. Tournament Selection requires more execution time while Roulette Wheel Selection allows the weaker chromosomes to be selected many times, so we choose Normalized Geometric Selection to choose the parents.

After parents being selected, the crossover operation will be done. We use arithmetic crossover (arithXover.m) function because it is specifically being used for floating point numbers and provides more than one crossover points. And we set four crossover points because our chromosome comprise of three alleles, one point crossover can not accommodate three alleles in one operation.

Mutation is done by setting mutation probability around 0.1 percent. In general mutation operations should not be done too often because the searching process will change into random search as the mutation probability getting higher.

## 5. APPLYING THE GENETIC ALGORITHM

There are several variables used as the standard to measure systems performance. In general, unit step input is used to test the systems, and the output signals is characterized by some *standard performance measures*: settling time, percent overshoot, error signal, rise time, peak time, and stability margin. All these measures are defined in time domain response.

Figure 3 below describes standard performance measures of a typical system driven by unit step input. Percent overshoot is defined as the point where the system response reaches the peak, in this case 53%. There are several criteria for settling time, for example 1% criterion, 2% criterion, and 5% criterion. Here we use 5% criterion settling time. And for the rise time, actually in general, is measured as the time needed by systems to reach from 0 to 100% of final value or from 10% to 90% of final value. But, for measurement simplicity, we use 0 - 95% criterion. Peak time is the point where the maximum value reached (overshoot) at 3.2 second. And error signal is the difference between the input signal magnitude and system response final magnitude. In this work, we use $G(s) = 1/s+1$, delay is in the range of 0.01 to 1 second. And because the systems are compensated by PID controller, the error signals are always zero. In addition to the five system standard performance measures described above, in Iterative Method and Ziegler-Nichols rule, we calculate the system performance indices described by equation (1) also. This is done because we want to compare it with the result of GA, which it's being optimized in the term of performance indices. Ideally, we can expect corresponding GA's performance indices should be always better than two tuning rules.

To calculate performance indices, we approximate the integral in equation (1) with addition (sigma) and 0.01 second sampling time and set the sigma upper limit with 15 second for all analyzed cases, no matter how quick it reaches convergence values.

## 6. SIMULATION RESULTS AND ANALYSIS

**Figure 3**. Standard performance Measures

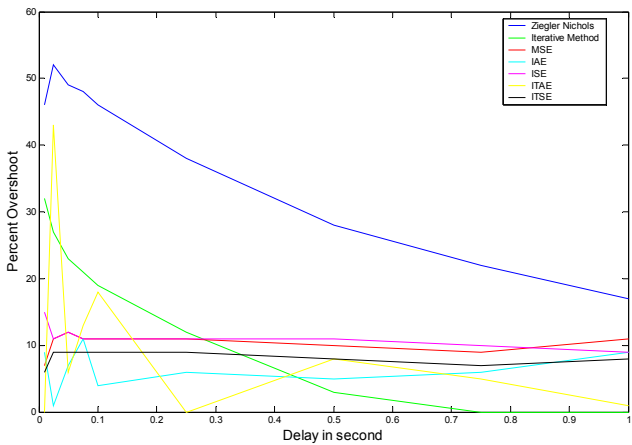

(a). The comparison of percent overshoots (PO).

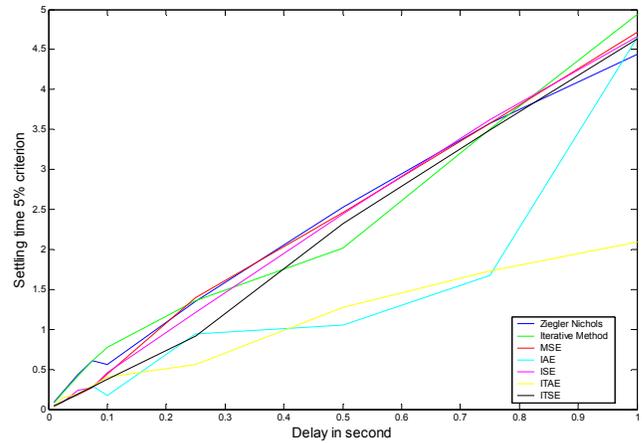

(b). The comparison of settling time (ST).

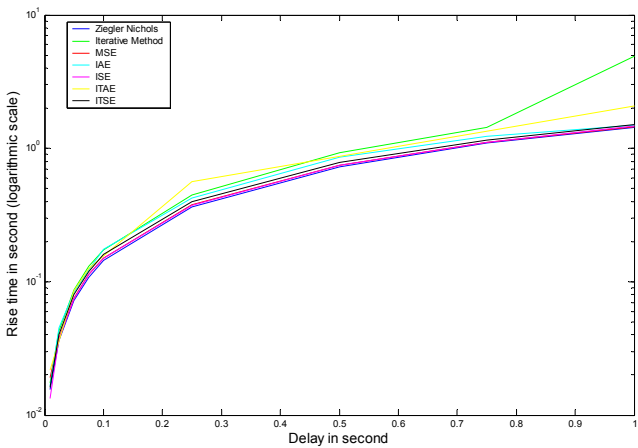

(c). The comparison of rise time (RT).

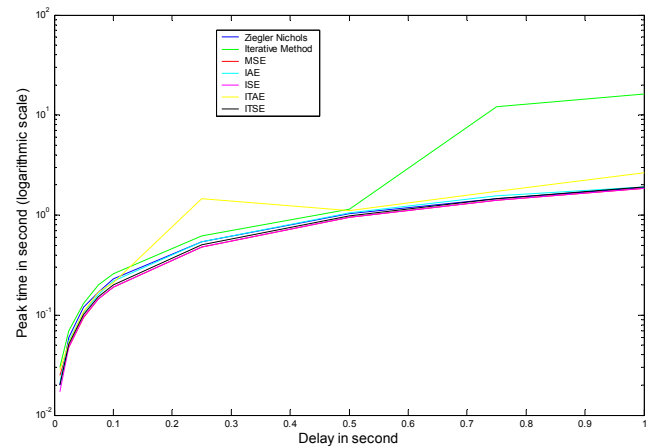

(d). The comparison of peak time (PT).

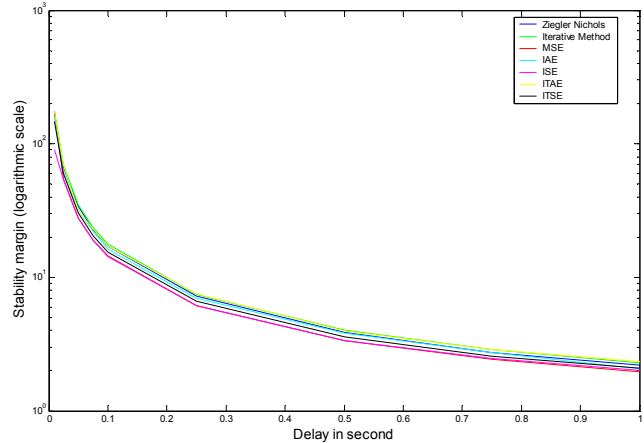

(e). The comparison of stability margin (SM).

Table 1. Average values of standard performance measures.

| Parameter | Ziegler-Nichols | Iterative Method | Optimized by MSE | Optimized by IAE | Optimized by ISE | Optimized by ITAE | Optimized by ITSE |
|---|---|---|---|---|---|---|---|
| %OV | 38% | 15% | 10% | 6% | 11% | 10% | 8% |
| ST5% | 1.53 | 1.54 | 1.47 | 1.01 | 1.45 | 0.745 | 1.37 |
| RT | 0.444 | 0.912 | 0.453 | 0.495 | 0.455 | 0.588 | 0.474 |
| PT | 0.613 | 3.43 | 0.576 | 0.622 | 0.576 | 0.836 | 0.597 |
| SM | 36.25 | 37.86 | 33.68 | 33.8 | 24.4 | 36.8 | 32 |

**6.1. Standard performance measures.**

*Percent Overshoot*

While figure 3(a) summarizes the values change of percent overshoots with respect to the time delay, table 1 gives its average values. Ziegler-Nichols rule gives the biggest value for all time delay, consequently its average value is the biggest also, 38%. Here the difference between two tuning methods and GA methods can be seen: while tuning methods have almost the same pattern, its value decreasing as the time delay growing bigger, except for small value of delay where Ziegler-Nichols rule gives increasing values, the GA methods give almost constant value, around 10% as long the time delay is not too small, except for case optimized by ITAE, where the percent overshoot value fluctuates and decreasing as time delay increasing. Table 1 shows that GA produces much better percent overshoot than two others tuning methods, especially if optimized by IAE criterion. So it can be concluded that GA can be used to optimized percent overshoot.

*Settling Time*

The value of settling time (5% criterion) all over the time delay is summarized by figure 3(b), where it can be seen that almost all methods, except for GA optimized by IAE and ITAE, fall under almost the same straight line with positive slope. It means that as the delay increasing, the settling time will increase linearly. In table 1 we can see that the settling time average value is not so different among all methods, in the range 1.37 second to 1.54 second, except for GA optimized by ITAE 0.745 second and IAE 1.01 second. However, because the others GA methods results are in agree with two tuning methods, we can't really differentiate these results. So it can be said that the settling time is not optimized by GA methods.

*Rise Time*

The third variable is rise time, which plotted in figure 3(c) using logarithmic scale in y axis. We can see strong pattern which all the results, except for Iterative Method, have almost the same value all over the time delay. And it should not be surprising if we get almost the same average value for all result, around 0.44 second to 0.59 second, except for Iterative Method, 0.912 second. Another interesting thing is, in general, almost in all range of time delay the curves keep their ranking unchanged, with order from the biggest value: Iterative Method, ITAE, IAE, ITSE, ISE, MSE, and Ziegler-Nichols rule.

From the average value (table 1), the best result is given by Ziegler-Nichols rule, 0.444 second and the worst one is Iterative Method 0.912 second, and all GA methods produce almost the same average value. But because GA results are not so different compared to Ziegler-Nichols rule, it can't be concluded that GA can optimize the rise time.

*Peak Time*

Almost the same pattern, as in rise time plots, is shown on the peak time plots on figure 3(d), except Iterative Method, in large delay range, tends to diverge, where all methods show almost the same values for all over time delay. But we must pay attention on GA optimized by ITAE because there is a range which its value is bigger than the others. Except for Iterative Method which is 3.43 second, all others methods produce peak time around 0.57 second to 0.83 second. The best values are given by GA optimized by MSE and ISE, 0.576 second. Like rise time, in general, almost in all range of time delay the curves keep their ranking unchanged, with order from the biggest value: Iterative Method, ITAE, IAE, Ziegler-Nichols, ITSE, MSE, and ISE. And because the GA methods produce peak time plots which only better than Iterative Method, not Ziegler-Nichols, the peak time can't be optimized by GA methods.

*Stability Margin*

The last standard performance measure is stability margin (figure 3(e)). Stability margin is the maximum gain that can be set before system response goes into sinusoidal cycle. In the simulations, this is done by simply increasing value of $K_c$ until sinusoidal cycle happens, and the stability margin of corresponding system is $K_c$ at sinusoidal cycle.

This is the first result that shows consistency all over time delay range, which all curves fall under almost the same line. So this is the strongest patterns, and because the stronger the pattern, the less the ability of GA methods to optimize corresponding performance measures, we can't use GA methods to optimize the stability margin. Conversely, the GA methods are likely to produce less stable systems.

Like rise time and peak time, in general, almost in all range of time delay the curves keep their ranking unchanged, with order from the biggest value: Iterative Method, ITAE, Ziegler-Nichols, IAE, ITSE, ISE, and MSE. But unlike settling time, rise time, and peak time, stability margin values reduce as the time delay increases.

**6.2. Performance Indices**

We differentiate the term standard performance measures and performance indices here, where standard performance measures have already been discussed above, performance indices are: MSE, IAE, ISE, ITAE, and ITSE. Table 2 below summarizes performance indices from simulation results. As expected, the average values of GA performance indices are always smaller than its corresponding Ziegler-Nichols and Iterative Method. Moreover Ziegler-Nichols rule produces smaller average performance indices values than Iterative Method does for all time delay values range.

**Table 2**. Performance indices of tuning methods and GA methods.

| Delay | Ziegler-Nichols | | | | | Iterative Method | | | | | Optimized by MSE | Optimized by IAE | Optimized by ISE | Optimized by ITAE | Optimized by ITSE |
|---|---|---|---|---|---|---|---|---|---|---|---|---|---|---|---|
| | MSE | IAE | ISE | ITAE | ITSE | MSE | IAE | ISE | ITAE | ITSE | | | | | |
| 0.01 | 0.000973 | 2.992285 | 1.461115 | 0.073894 | 0.020615 | 0.000927 | 2.730597 | 1.391095 | 0.061045 | 0.014833 | 0.00081 | 1.623776 | 0.840777 | 0.011395 | 0.005388 |
| 0.025 | 0.002626 | 6.927413 | 3.942144 | 0.432876 | 0.124957 | 0.002313 | 6.300772 | 3.471926 | 0.391845 | 0.089213 | 0.001884 | 3.701027 | 2.827366 | 0.218351 | 0.038659 |
| 0.05 | 0.004988 | 12.98842 | 7.487622 | 1.580628 | 0.452425 | 0.004376 | 11.6831 | 6.567769 | 1.452929 | 0.318439 | 0.003708 | 7.09663 | 5.57464 | 0.301924 | 0.155625 |
| 0.075 | 0.007197 | 18.53609 | 10.80309 | 3.249557 | 0.929874 | 0.006334 | 16.58511 | 9.507315 | 2.979748 | 0.642704 | 0.005508 | 10.75044 | 8.266998 | 0.754851 | 0.350289 |
| 0.1 | 0.009284 | 23.69959 | 13.93484 | 5.304512 | 1.517501 | 0.008226 | 20.99624 | 12.34746 | 4.801577 | 1.035597 | 0.0073 | 13.86607 | 10.95693 | 1.526731 | 0.622276 |
| 0.25 | 0.020422 | 48.31448 | 30.65349 | 20.82159 | 6.345217 | 0.019092 | 40.69953 | 28.65714 | 15.58065 | 4.507248 | 0.017983 | 34.18455 | 26.99306 | 7.69427 | 3.854827 |
| 0.5 | 0.037119 | 76.26794 | 55.71523 | 45.56416 | 17.73724 | 0.037603 | 71.60299 | 56.44149 | 37.68745 | 16.79849 | 0.035612 | 67.45696 | 53.45713 | 30.00848 | 15.15463 |
| 0.75 | 0.05416 | 108.6372 | 81.29453 | 91.98153 | 36.10692 | 0.057392 | 123.8553 | 86.14606 | 146.8766 | 42.31687 | 0.053071 | 100.0041 | 79.65463 | 63.58332 | 33.54903 |
| 1 | 0.072117 | 146.0676 | 108.247 | 174.114 | 65.18471 | 0.077665 | 176.087 | 116.5754 | 298.2008 | 81.12556 | 0.070379 | 134.0671 | 105.6145 | 111.3298 | 58.69364 |
| Av. | 0.02321 | 49.38123 | 34.83767 | 38.12475 | 14.26883 | 0.02377 | 52.28229 | 35.6784 | 56.44808 | 16.31655 | 0.021806 | 41.41674 | 32.68734 | 23.93657 | 12.4916 |

Although it can be seen that MSE has the smallest average values for all three methods, and IAE has the largest average values for Ziegler-Nichols and GA methods, and only second in Iterative Method, it doesn't imply that one must use MSE and must avoid using IAE as a objective function in GA because these performance indices, as shown by equations (1), have different definitions and cannot be compared. Moreover, as shown in table 1, GA method optimized by MSE doesn't produce the best results for all analyzed standard performance measures, only for peak time. To get more insight about the comparison among these methods, we plot values change of each performance indices with respect to time delay below.

**Figure 4**. Performance Indices

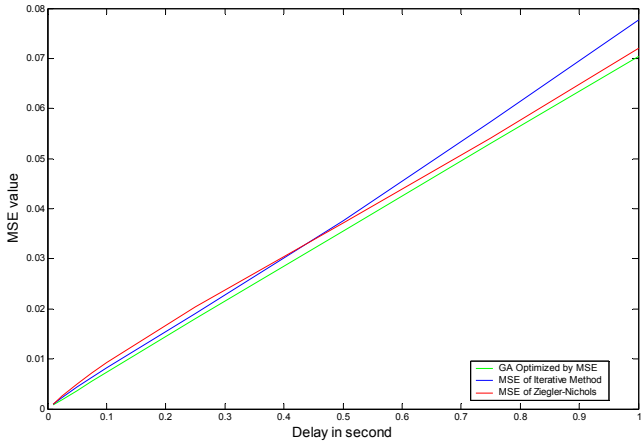

(a). The comparison of MSE values.

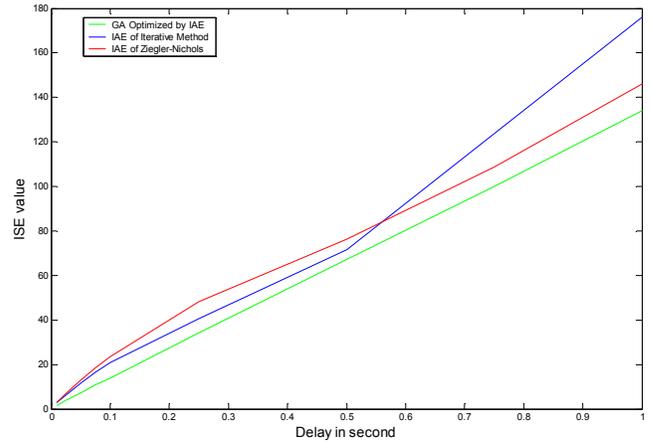

(b). The comparison of IAE values.

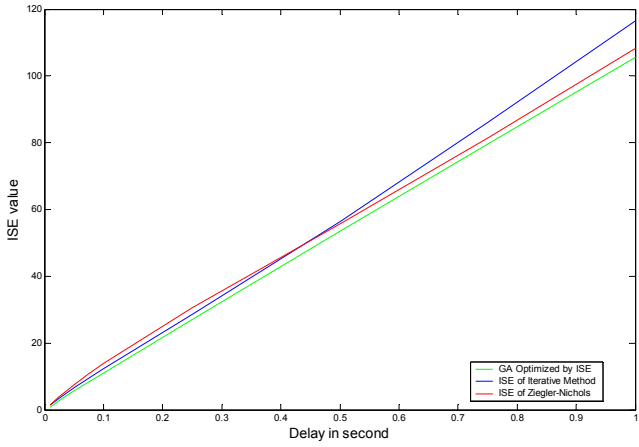

(c). The comparison of ISE values.

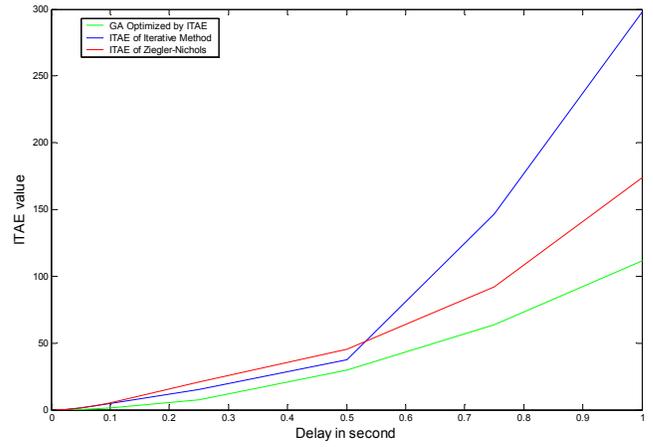

(d). The comparison of ITAE values.

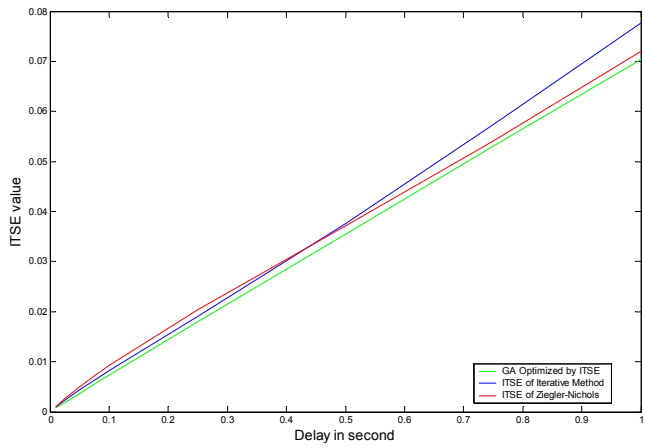

(e). The comparison of ITSE values

All five figures above confidently show that GA method gives the smallest values of all performance indices analyzed for all range of time delay. So not only in average values, but also for all measured values does GA method produce the smallest corresponding performance indices. However, the differences between GA method and two tuning methods results, except for ITAE objective function where the differences increase as the time delay increases, are not impressive enough to come into conclusion that GA method is much better than two others methods in minimizing error criteria. Besides, we must consider the convergence problem arises in applying GA, which in this work the experiments don't always come into desired solutions. Even though we set the value bounds based on the previous results from two tuning methods, it only improves the probability that the simulations come into convergence results (from 45 cases, there two times failed to reach convergence results).

## 7. CONCLUSIONS

1. Genetic Algorithm applied in PID controller improves FOLPD transient response compared to two tuning methods. This is shown by average percent overshoot reduction, more than 70% and 30% with respect to the Ziegler-Nichols rule and Iterative Method, while keep the rise time and peak time almost unchanged and improves the settling time. However, there is payoff in the stability margin which reduces slightly compared to two tuning methods.
2. The average values of GA performance indices, as expected, are always smaller than its corresponding Ziegler-Nichols and Iterative Method. Moreover Ziegler-Nichols rule produces smaller average performance indices values than Iterative Method does for all time delay values range. However, the differences between GA method and two tuning methods results, except for ITAE objective function where the differences increase as the time delay increases, are not impressive enough to come into conclusion that GA method is much better than two others methods in minimizing error criteria.
3. There are convergence problems that arise in applying GA, which in this work the experiments don't always come into desired solutions. Even though we set the value bounds based on the previous results from two tuning methods, it only improves the probability that the simulations come into convergence results. Moreover, the value bounds setting based on tuning methods results discards the possibility to find optimum results from others value ranges.

*Profile*

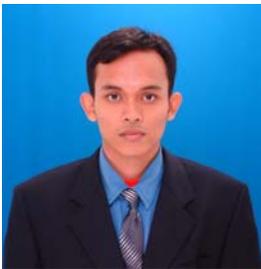

**Andri Mirzal** is a student at Graduate School of Information Science and Technology, Hokkaido University. He obtained his Bachelor degree (Honor) in Electrical Engineering majoring in Control Engineering from Bandung Institute of Technology, Indonesia, 2003. His research interests are in optimization methods, control systems, and network theories. He is now carrying out his research program under supervision of Professor Masashi Furukawa, Autonomous Systems Engineering Laboratory, Graduate School of Information Science and Technology, Hokkaido University.

**Shinichiro Yoshii**

**Masashi Furukawa**